# Energy Balanced Two-level Clustering for Large-scale Wireless Sensor Networks based on the Gravitational Search Algorithm

Basilis Mamalis[1]

University of West Attica
Agiou Spyridonos, 12243, Egaleo
Athens, GREECE

Marios Perlitis[2]

Democritus University of Thrace
University Campus, 69100
Komotini, GREECE

*Abstract*—**Organizing sensor nodes in clusters is an effective method for energy preservation in a Wireless Sensor Network (WSN). Throughout this research work we present a novel hybrid clustering scheme that combines a typical gradient clustering protocol with an evolutionary optimization method that is mainly based on the Gravitational Search Algorithm (GSA). The proposed scheme aims at improved performance over large in size networks, where classical schemes in most cases lead to non-efficient solutions. It first creates suitably balanced multihop clusters, in which the sensors energy gets larger as coming closer to the cluster head (CH). In the next phase of the proposed scheme a suitable protocol based on the GSA runs to associate sets of cluster heads to specific gateway nodes for the eventual relaying of data to the base station (BS). The fitness function was appropriately chosen considering both the distance from the cluster heads to the gateway nodes and the remaining energy of the gateway nodes, and it was further optimized in order to gain more accurate results for large instances. Extended experimental measurements demonstrate the efficiency and scalability of the presented approach over very large WSNs, as well as its superiority over other known clustering approaches presented in the literature.**

## I. INTRODUCTION

Grouping sensors in clusters is an effective method for saving energy in large-scale WSNs [1]. Considering such a WSN, there are several sets of sensors called clusters and each one of them has a leader called 'cluster head'. The sensor nodes (after sensing the field) send the sensed data to the CH, and then the CH (after collecting the data) forwards them to the BS. A study in WSN clustering is given in [1]. Furthermore, a lot of scientists have adopted the use of a specific kind of nodes called 'gateway-nodes' which operate similarly to the normal sensor nodes, however they are usually equipped with more energy and communication capabilities and they cost more [2]. The gateway nodes finally behave as CH in the WSN, and forward the gathered data to the BS. Actually the scope of this idea is to create a stronger group of CH than in typical networks with cluster organization. On the other hand, their operation is still based on batteries, so they need to preserve their energy adequately while the network operates.

Let's suppose we have $m$ gateway nodes and $n$ sensors in the network, and there are $k$ nodes in the range of the gateway nodes. The complexity for associating the $n$ sensors to the $m$ gateway nodes is naturally exponential. So, typical techniques can't normally lead to effective solutions that scale well. A lot of evolutionary methods could be applied to give a good approximate solution. The GSA [3] has been recently reported as a quite valuable such technique, driven by nature, which can manage effectively problems in hard computational form. Also, the particle swarm optimization technique and the ant colony method have been thoroughly studied [1].

A large number of research works can be found in the bibliography [1, 4-17, 22-32] following clustering schemes in WSNs (either centralized or distributed). The LEACH protocol [8] is the most known and one of the most effective. In this protocol the cluster-heads are chosen on the basis of a specified probability. The disadvantage of this protocol is that a sensor node of less energy could be chosen as a cluster head, thus limiting the network operation. PEGASIS [9] was proposed as an efficient alternative, but it can't lead to good scalability because of several constraints. A number of other schemes adopting clustering have further been met, trying to achieve sufficiently balanced depleting of energy and thus improving the network operation. Many of them [4,22-23,31-32] aim at solving the '*hot-spot*' problem and achieving ideal distribution in the depletion of energy for all the nodes, whereas others [28-30] try to gain efficiency and scalable secure behavior when dealing with large-scale networks and deployment areas.

Focusing on the schemes based on 'gateways', in [13] the authors propose a scheme of satisfactory balance, named LBC. In LBC the clusters are formed quite efficiently, but the algorithm doesn't take in account neither the distance between the sensor nodes and the gateway nodes nor the remaining energy of the gateway nodes. In [14] the authors propose an approach named GLBCA applying BFS, but when large-scale WSNs are considered non-satisfactory execution times are observed. In [15] the authors present an approach based on GA-clustering in order to elect a set of cluster heads among the regular sensors. In [16] the authors propose a scheme based on PSO by taking in account the energy and the distance within the clusters as main criteria, but on the other hand they don't consider the energy and the distance of the gateway nodes.

In [17], a clustering scheme of high efficiency is presented built over the GSA approach, in which the fitness function is properly selected considering not only the distance among the sensor nodes and the gateways/BS, but also the remaining energy of the gateway nodes. The eventually proposed scheme, named GSA-EEC, is evaluated based on several metrics, and it is proved to be very effective and performing better than most of the related approaches [13-16]. On the other hand, when large-scale networks are considered, these methods in most cases can't offer satisfactory results because of various problems met in the case of very high number of dimensions in the search space. More concretely, the GSA method in such cases is shown to have limited stability as well as low levels of accuracy due to the increased possibility to be trapped into local optimum solutions; additionally non-efficient execution times are observed. The above shortcomings restrict the worth of using the GSA technique in very large networks. The GSA approach has also been used in WSNs for nodes localization and relay/sink nodes placement, as shown in [18-21].

Trying to overcome the disadvantages referred in the previous paragraph with regard to the use of the GSA-oriented scheme [17] on large networks, we've designed a novel hierarchical clustering protocol that consists of two levels, and combines adequately the above GSA-oriented scheme with a classical gradient clustering scheme, similar to [22-23] and suitably modified in order to generate less number of clusters and fit better in the energy balancing needs of our proposed total scheme. The presented approach first constructs multihop clusters of evenly distributed energy reserves, in which the sensor nodes energy gets larger as coming closer to the cluster head. Thus, since the sensors near to the cluster head are more exhausted because of data forwarding, keeping the residual energy of those sensors high guarantees the prolonged and seamless operation of the WSN. Next, a suitable protocol based on the GSA runs to associate sets of cluster heads to specific gateway nodes for the eventual forwarding of data to the base station. Extended simulated experiments are presented to show the high efficiency and scalability of the proposed combined clustering scheme over very large WSNs, as well as its superiority over other clustering approaches of the literature [26-27,30-31]. Further experiments have been done to show the suitability of the proposed new fitness function against the one referred in [17], as well as the worth of using the proposed modified first-level clustering algorithm against the use of other alternatives (i.e. [22-23]).

The primary goal of our work is to build a clustering protocol able to achieve highly efficient behavior in *very large-scale* WSNs (i.e. not only for hundreds or one-two thousands of sensor nodes but also for 5000-10000 and even more sensor nodes; as it will be the case in future IoT/IoE applications). The existing clustering protocols unfortunately suffer from several shortcomings that lead to performance degradation in such *very large scale* WSNs, either relevant to the *hot spot* problem or to the fact that in order to face the *hot spot* problem they tend to generate quite large number of CHs and/or multihop routing overhead when we have large numbers of sensors (making their further use problematic, e.g. for data gathering etc.), and generally lead to not sufficiently scalable solutions. As a consequence a proper solution could be given by combining appropriately clustering protocols with specific features into corresponding two-level hierarchical schemes trying to gain efficiency and extent the scalability of the total scheme.

The main contribution of our work includes the following:

- The appropriate modification of the gradient-based clustering protocol presented in [22-23] in order to generate less number of clusters, achieve better execution times, and generally fit better in the balancing needs and the total behavior required as a first-level (only) clustering protocol.

- The appropriate modification of the GSA-based clustering protocol presented in [17], mainly with regard to its fitness function in order to have more accurate results (with regard to energy balancing), as well as with respect to the total behavior required as a second-level clustering protocol over a number of CH nodes.

- The integration of the above two basic clustering components (first-level and second-level) into a single combined (two-level) clustering approach that achieves both high energy efficiency and balancing, and also preserves its high scalability for very large WSN instances and relevant deployment areas due to its two-level nature.

- The extended evaluation (via carefully designed simulation experiments with Castalia simulator [25]) of both the two basic clustering components and the total proposed (two-level) clustering scheme, in order to demonstrate its high efficiency and scalability, in terms of energy consumption and network lifetime over very large WSNs, as well as its superiority over other relevant approaches.

The remaining text is organized as follows. In Section II some background is given with respect to the GSA. The proposed clustering scheme is described in Section III. In Section IV our simulated experiments are presented and discussed, and in Section V the main conclusions are stated.

## II. GRAVITATIONAL SEARCH ALGORITHM

The description of the GSA technique in details is given in [3]. Let's assume we have a group of agents, $N_A$. Every agent is expected to give a part of the solution. The location of agent $A_i$, $1 \le i \le N_A$ in dimension $d$ is $x_i^d$ whereas its velocity is $v_i^d$, $1 \le d \le D$. Every agent has the same dimension. Every agent is evaluated to verify the suitability of the result based on a specific fitness function. Let the $i^{th}$ agent's location be represented as $X_i = (x_i^1, x_i^2, ..., x_i^D)$. The following expression gives the force applied on the $i^{th}$ agent by the $j^{th}$ agent.

$$F_{ij}^d(t) = G(t) \frac{M_{pi}(t) \times M_{aj}(t)}{R_{ij}(t)} (x_j^d(t) - x_i^d(t))$$

$M_{pi}$ represents the passive mass of the $i^{th}$ agent, whereas $M_{aj}$ stands as the corresponding active mass of the $j^{th}$ agent, and $\alpha$ is a constant. $G(t)$ equals to $G_0(t_0/t_{max})^\beta$, in which $G_0$ represents also a constant. $R_{ij}(t)$ means the Euclidean distance from agent $i$ to agent $j$. The overall force applied by the group of agents over the $i^{th}$ agent in dimension $d$ is as follows.

$$F_i^d(t) = \sum_{j=1, j\neq i}^{Na} rand_j \times F_{ij}^d(t)$$

The inertial mass as well as the gravitational mass is estimated by evaluating the fitness function. The heavier the mass of an agent the more efficient the agent is.

$$m_i(t) = \frac{fit_i(t) - worst(t)}{best(t) - worst(t)} + \varepsilon$$

In the above, $\varepsilon$ is a limited constant, $fit_i(t)$ represents the fitness of agent $i$, whereas $worst(t) / best(t)$ are as next.

$$best(t) = \min_{j\in\{1...Na\}} fit_j(t) \quad worst(t) = \max_{j\in\{1...Na\}} fit_j(t)$$

$$M_i(t) = \frac{m_i(t)}{\sum_{j=1}^{Na} m_j(t)}$$

Further, let's assume $M_{ai}$, $M_{pi}$, $M_{ii}$ and $M_i$ are all equal to each other. Based on the $2^{nd}$ law of Newton, we have the next.

$$a_i^d(t) = \frac{F_i^d(t)}{M_{ii}(t)}$$

In the above, the inertial mass of agent $i$ is given by $M_{ii}$, whereas the acceleration of agent $i$ is given by and $a_i^d(t)$.

### III. PROPOSED CLUSTERING APPROACH

As previously stated, the proposed approach has been developed on the basis of a hierarchical clustering protocol of two levels, named GC-GSA. First, the sensors and the gateway nodes are spread at random over a large deployment area. The deployed sensors / gateways are supposed to be static; no mobility is supported. Then, a two-phase network operation begins. In the first phase the necessary bootstrap and cluster formation procedures are completed. Initially, proper identities are assigned to all the sensors by the base station. Then, both the sensors and the gateway-nodes send their identities and other info to their neighbors using mac-level protocol. The gateway-nodes are finally informed with the identities of their neighbor sensors. Next, every gateway-node informs the base station with the collected info to complete the setup. The complete clustering routine is the run and all the sensors nodes get the necessary info with regard to their CH identity. In the second phase the network starts to operate steadily; the sensed data are gathered by the cluster heads and then by the gateway-nodes, and they finally directed to the base station.

### A. First Level Energy-Balanced Clustering

With regard to the initial clustering, a suitable protocol based on multiple hops communication is adopted, aiming at building a robust hierarchy with controlled delay and sufficient coverage with respect to the secondary clustering (e.g. the size of the clusters and the positions of the selected cluster-heads). Further, the algorithm of [4] is primarily followed as the firstly applied clustering routine, whereas a proper adaptation has been incorporated aiming at inheriting the main advantages of approach described in [22, 23]. The remaining energy of the sensors is chosen as the basic factor while building the clusters, so a sufficiently balanced cluster hierarchy is built, the energy-hole problem (near to the cluster-heads) is effectively handled, and eventually the operation of the network is prolonged. A detailed description of the relevant procedure is given in [33].

From a qualitative point of view our modified clustering protocol is quite similar to the watershed-based clustering protocol presented in [22-23], with the main difference that it doesn't take in account separately (as separate clusters) the sets of sensors having almost equal to each other remaining energy. In our modified algorithm all the sensors are forced to follow a parent with larger residual energy (in the way described above) even if the residual energy is quite close (approximately the same) to the residual energy of the elected parent. As a result it leads to less number of CHs (which is a crucial factor in the context of the present gateways-based WSN environment), whereas also the execution of the cluster formation procedure is quite faster. On the other hand (validated also clearly by the corresponding simulations given in section 4.4), no actual 'loss' in the energy balancing performance is observed with respect to the present gateways-based WSN problem, if compared to the probable use of the watershed-based algorithm as the first-level clustering protocol in the same problem. Actually the slight loss that one would expect is counterbalanced by the energy savings due to the less number of cluster-heads and consequently the less heavy communication among the cluster-heads and the gateways, thus finally leading to slightly better overall performance (specifically when considering large and very large numbers of sensors).

On the basis of the above considerations, we have followed the use of the proposed modified algorithm as the first-level clustering protocol (as opposed to the use of the watershed-based algorithm of [22-23] or the use of the much simpler but weaker protocol of [4]), mainly because it has the potential to perform and scale better for large / huge numbers of sensor nodes and relevant deployment areas, in the context of this problem. Instead, in the context of other WSN environments and relevant applications (e.g. data gathering applications with mobile sinks where a number of sinks can be used to visit all the CHs, as in [22-23]) the performance of the watershed-based clustering protocol should be the preferable one.

### B. Second Level GSA-based Clustering

Considering the next (second) level of building the cluster hierarchy, the GSA oriented protocol of [17] has been suitably modified and applied over the total set of cluster-heads elected in the first level (which is of relatively small size - i.e. avoiding the use over the complete WSN). The adopted GSA oriented scheme runs to associate sets of cluster-heads to specific gateway nodes for the eventual relaying of data to the base station. The initially chosen fitness function was further optimized and modified in order to keep the energy balanced over the gateway-nodes too, and gain more accurate results for large instances. The initialization of the basic group of agents ($N_A$) first takes place (note that each agent potentially stands as a part of the solution). More concretely, in the context of this stage of clustering the agents represent the associations of cluster-heads to corresponding gateway-nodes. If we assume that $A_i$ stands as the agent $i$, every item $x_i^d(t)$ associates the relevant cluster-head to some gateway and $1 \leq i \leq N_A$, whereas also $1 \leq d \leq D$ (note that $D$ equals to $c$). So, each agent may be denoted in the following form [17, 33], whereas our modified GSA protocol should than run as follows.

$$A_i = [x_i^1(t), x_i^2(t), x_i^3(t), \ldots, x_i^D(t)]$$

### The Second-level GSA-based Clustering Algorithm

**Input:**

–  Group of CHs: $H = \{h_1, h_2, h_3, \ldots, h_c\}$

–  Set of gateway nodes: $G = \{g_1, g_2, g_3, \ldots, g_m\}$

–  Initial group of agents with size equal to $N_A$

–  Agent's dimensions = # of CHs = $c$

**Output:**

The optimal CHs associations to gateway-nodes

**Description of the algorithm:**

Agent $A_i$ is initialized, $\forall i, 1 <= i <= N_A$

The mapping function is defined (for every $h_d$ to a $g_k$)

***do***  /* initially assume that $t=0$ */

 **for** $i=1$ to $N_A$

 Fitness $(A_i)$ is computed

 The *best/worst* fitness values are updated of all agents

 $M_i(t), a_i^d(t)$ are computed of each agent of the system

 The velocity and the position of $A_i$ are updated

 **endfor**

**while** the criteria for termination are not satisfied

### C. Choosing the Fitness Function

The definition of fitness function ($f$) has to be appropriately specified considering not only (i) the residual energy of the gateway-nodes, but also (ii) the distance between the cluster-head and the gateway-node as well as between the latter and the base station. The gateway-nodes of high remaining energy reserves should be elected. Thus, the energy consumption is suitably balanced and the lifetime of the network is prolonged. Moreover they should be the ones having the less distance too. These requirements could be further described as follows ($E_{gj}$ stands as the remaining energy of $g_j$, and $l_{ij}$ equals to the number of gateway-nodes in the neighborhood of $h_i$).

| Fitness Function #1: | |
|---|---|
| Object. 1: *Maximize f1 =* | $\sum_{i=1}^{c} E_{gj}$ |
| Object. 2: *Minimize f2 =* | $\sum_{i=1}^{c} (d(h_i, g_i) + d(g_i, BS))$ |
| Total Object.: *Fitness f =* | $a \times \dfrac{1}{f1} + \beta \times f2$ |
| Get $f$ minimized where $\alpha + \beta = 1$ | |

Several other functions denoting fitness (relevant fractions with a factor that minimizes the distance and a factor that maximizes the energy in the right place) could similarly be used and evaluated. Instead, the original fitness function given in [17] was initially adopted, aiming at having the same basis for comparing both approaches in our experimental evaluation. In our experiments $\alpha$ and $\beta$ have been adequately defined to give us the best performance evaluation measurements in every instance. Beyond this fitness function (which was applied and tested as our basic selection due to comparison reasons), we've also chosen a suitable alternative function which is presented as follows (fitness function #2).

| Fitness Function #2: | |
|---|---|
| Object. 1: *Maximize f1 =* | $\sum_{i=1}^{c} E_{gj}$ |
| Object. 2: *Minimize f2 =* | $\sum_{i=1}^{c} (d(h_i, g_i) + d(g_i, BS))$ |
| Total Object.: Minimize $f =$ | $\dfrac{\beta \times f2 + t_1}{\alpha \times f1 + t_2}$ |

Note that $\alpha$, $\beta$ and $t_1, t_2$ are residual-energy/distance dependent and independent constants respectively, whereas the final value of the function is restricted adequately between 0 and 1 to normalize and optimize the result. The new proposed fitness function not only balances the weight of the two main factors (residual energy and distance) in the final computation, but also allows the designer to normalize (through $\alpha, \beta$) conveniently the unpredictable (non-canonical) gaps caused by the potentially different measure units. It gives also the flexibility to take in account (through $t_1, t_2$) during the optimization process other significant parameters too (like data rates, transmission range, initial energy and energy consumption rates etc.). As it is shown in subsection IV.B the new fitness function leads both algorithms to better / more accurate execution behavior.

### D. Discussion and Extensions

As it comes out of the literature, all the existing clustering protocols unfortunately suffer from several shortcomings that lead to performance degradation in very large scale WSNs, either relevant to the *hot spot* problem or to the fact that in order to face the hot spot problem they tend to generate quite large number of CHs and/or multi-hop routing overhead as the number of sensors increases, and generally lead to not sufficiently scalable solutions. As a consequence a proper solution could be given by combining appropriately clustering protocols with specific features into corresponding *two-level* hierarchical schemes trying to gain efficiency and extent the scalability of the total scheme.

In the above context, with the use of the GSA oriented scheme as the upper level cluster formation routine, we naturally get over the related disadvantages, due to the fact that it operates on a quite restricted number of nodes/cluster-heads. Thus, we may gain from its efficient behavior over such WSNs (of restricted size), and finally conclude to an outstanding performance gains if we use it in large / huge WSNs together with a lower level cluster formation scheme of similar efficiency (considering the balance of energy consumption), like the protocol referred as the basis of our approach.

Specifically, as it is also referred in section 4, the GSA oriented scheme behaves with great efficiency when applied over a set of e.g. 500 sensors. As a consequence, our combined approach may behave similarly over e.g. up to 500 cluster-heads, thus having the potential to lead to excessive total performance for large / huge networks of 5000-10000 and even more sensors.

On the contrary, one should also note that the proposed gradient-oriented algorithm applied for the initial clusters formation isn't suitable enough to cluster effectively the full large-scale WSN (despite of the fact that such cluster formation algorithms are considered quite efficient for this kind of WSNs too [22-23]). The basic disadvantage of such algorithms [4,22-23] relies on the fact that in many times they conclude to excessive numbers of cluster-heads when very large numbers of sensor nodes are deployed in the field. As a consequence the task of relaying of the data to the base station (or gathering with the use of a mobile sink) gets harder and completes in non-satisfactory total times. The ideal balancing of energy, preserving the number of cluster-heads in acceptable levels, is very hard to be achieved without applying a multi-level clustering procedure. Thus, by using additional gateway-nodes in the upper level, a hybrid scheme of great efficiency may naturally be built. Finally, it should be noted that the presented hybrid cluster formation procedure could be efficiently combined and integrated (in a straightforward manner - e.g. like in [21-24]) with one or more (mobile or not) sinks / collectors, thus forming a robust and flexible data gathering solution for very large-scale WSN application environments.

For example, following an approach similar to the one of [21] or [22-23] and considering a very large WSN where the number of gateways is also large enough, one should efficiently drive a set of one or more mobile sinks (or place a corresponding set of stationary sinks) to specific optimal (with respect to the locations of the gateways) route/locations to gather the data collected by the gateway-nodes.

## IV. SIMULATION RESULTS

Next, our relevant simulation experiments are presented, in which the proposed approach is shown to achieve high efficiency and scalability. Also our approach is compared to the original GSA oriented scheme given in [17] without any modifications. Specific sets of experiments have also been executed to show the suitability of the newly designed fitness function against the one referred in [17], as well as the worth of using the proposed modified first-level clustering algorithm against the use of the watershed-based one of [22-23]. The simulation experiments have been completed with the use of Castalia (a WSN simulator built on the basis of OMNeT++ [25]), and they focus on measuring the corresponding in each case protocol's efficiency (measuring the consumption of energy and lifetime of the WSN), over very large networks. In order to have a right comparison, the radio model used is the same as in [8, 17]. The consumption of energy for every node while sending a packet of $l$ bits equals to the following:

$$E_{TX}(l,d) = \begin{cases} l \times E_{elec} + l \times \varepsilon_{fs} \times d^2, & \text{if } d < d_0 \\ l \times E_{elec} + l \times \varepsilon_{mp} \times d^4, & \text{if } d \geq d_0 \end{cases}$$

The energy consumption for transmitting or receiving one bit is $E_{elec}$, whereas the energy for amplification is based on the relevant model of the transmitting circuit, and $d_0$ stands as the limit for the maximum possible distance of transmission. Respectively, the energy consumption by the receiver for a packet of $l$ bits equals to the following:

$$E_{RX}(l) = l \times E_{elec}$$

### A. Comparing to the Native GSA-based Approach

Trying to compare the proposed hybrid scheme to the original GSA oriented scheme of [17], specific experiments have been run, considering different numbers of sensor nodes ($n$ equal to 500, 1500, 2500), deployed at random (together with gateway-nodes) in a grid area with its side taking values between 200m and 500m (i.e. deployment areas of 200x200m$^2$, 360x360m$^2$ and 480x480m$^2$). The two first parameter values (500, 200x200m$^2$) stand as the main setup size chosen in the experimental evaluation of [17]. Based on the remaining two setup parameter values an approximate scale of 3x and 5x is simulated with respect to the experiments of [17]. With regard to the network setup factors considered in our simulations (transmission range, initial energy, etc. for both the sensors and the gateways) we have used the same values as given in the experiments of [17]; for comparison reasons too. The most important parameter values are summarized in Table I. Finally, it must be noted that for both algorithms the setup procedure is adjusted adequately to have equivalent behaviour with respect to the acceptable limit of coverage of the field where the sensors are deployed. The corresponding results are given in Fig. 1-4. By observing Fig. 1-3 one can easily conclude that considering the consumption of energy for each node, our proposed approach (GC-GSA) is definitely superior to the native GSA-EEC approach when n is equal to 1500 and 2500. On the other hand, it achieves almost the same performance (slightly worse) when n is equal to 500 nodes.

Specifically, for n equal to 1500, GC-GSA is shown to have a decrement of 15% with respect to the energy consumed on average, and this decrement becomes equal to 26% for n equal to 2500 sensor nodes. Instead, GSA-EEC has an almost ideal behaviour when n is equal to 500 sensor nodes, and its behaviour becomes gradually worse and worse as the WSN increases in size (which is a normal expectation because of the disadvantages of GSA technique, mainly when the number of dimensions is large). Moreover, Fig. 4 presents the lifetime of the network (which is measured as the life duration of the sensor node that dies first) with respect to the various test scenarios addressed above. One may easily observe that the GC-GSA approach leads to significant increments of the WSN lifetime when considering large networks (raising up to 30% - and even more - when n is equal to 2500 nodes). Also, the GSA-EEC scheme has a little better performance when n is equal to 500 nodes. It must also be noted that the disadvantages of GSA-EEC have a direct impact on the WSN lifetime, more important even than the energy consumed in average, because the energy consumption variance increases significantly too.

TABLE. I. PARAMETER VALUES FOR SIMULATION EXPERIMENTS

| Parameter | Value |
|---|---|
| Sensor energy | 1J |
| Gateway energy | 5J |
| $E_{elec}$ (bit energy dissipation) | 50nJ/bit |
| $\varepsilon_{mp}$ (multipath model amplification energy) | 10pJ/bit/m$^2$ |
| $\varepsilon_{fs}$ (free space model amplification energy) | 0.0013pJ/bit/m$^4$ |
| $d_{max}$ (sensor max. transmission range) | 100m |
| $d_0$ (sensor threshold distance) | 86m |
| $R_{max}$ (gateway max. transmission range) | 150m |
| $R_0$ (gateway threshold distance) | 129m |
| Length of one data packet | 4000bits |
| Size of one message | 500bits |

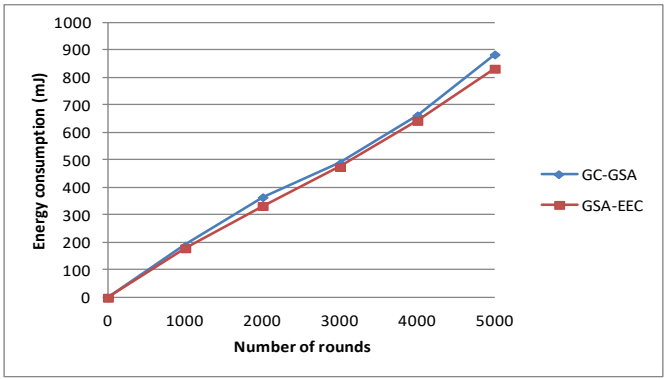

Fig. 1. Average Energy Consumption for n=500.

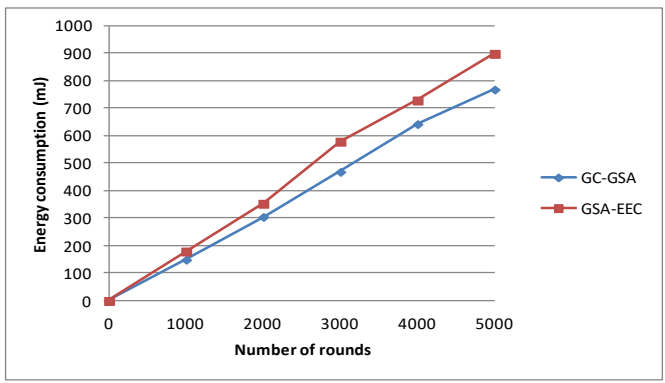

Fig. 2. Average Energy Consumption for n=1500.

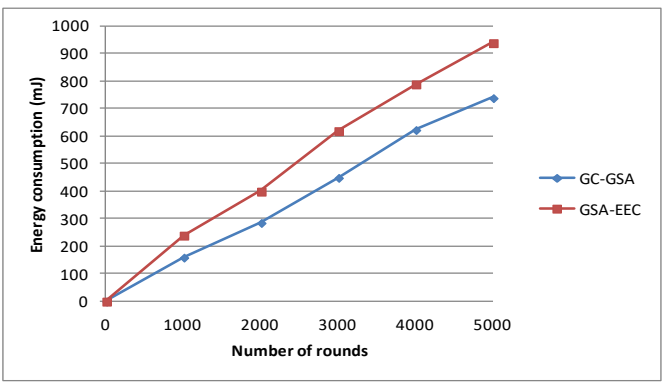

Fig. 3. Average Energy Consumption for n=2500.

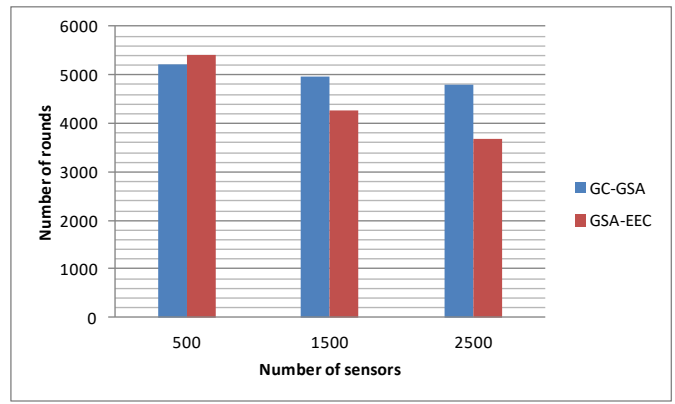

Fig. 4. Network Lifetime for Varying 'n'.

### B. Evaluating the New Proposed Fitness Function

In the next set of experiments (summarized in Fig. 5 and 6) we present the improvements introduced due to the use of our new proposed fitness function (as described and discussed in subsection III.C). The corresponding measurements have been taken for n=2500 sensors over a 480x480m$^2$ terrain and present the energy consumed in average and the lifetime of the network for both the GC-GSA and GSA-EEC algorithms, for each one of the two fitness functions (the previously chosen in [17] - FF #1 - and the new proposed one here - FF #2).

As shown in Fig. 5, a significant improvement in the average energy consumption of the GC-GSA algorithm is achieved with the use of FF #2, which ranges form 5.5% to 9.5% depending on the number of rounds. Similarly, the average energy consumption of the GSA-EEC algorithm is also improved with the use of FF #2 (a decrease ranging from 3.5% to 5.5%); it remains however substantially worse than the one of the GC-GSA algorithm.

The improvements observed for the GSA-EEC algorithm are relatively smaller than the ones of the GC-GSA algorithm, mainly due to the fact that the performance of the GSA-EEC algorithm is influenced dramatically (in any case and more or less independently to the form of the fitness function) from the shortcomings met for large number of sensors (GSA dimensions etc. as it has been discussed earlier in Section 1). Furthermore, in Fig. 6 the corresponding improvements in network lifetime (due to the use of FF #2) are given for both algorithms.

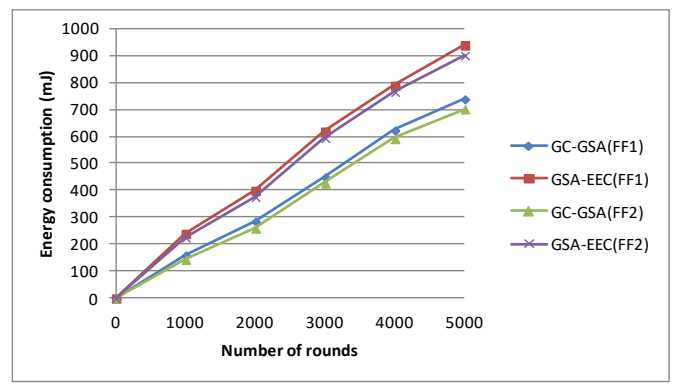

Fig. 5. Average Energy Consumption for Fitness Functions #1 and #2.

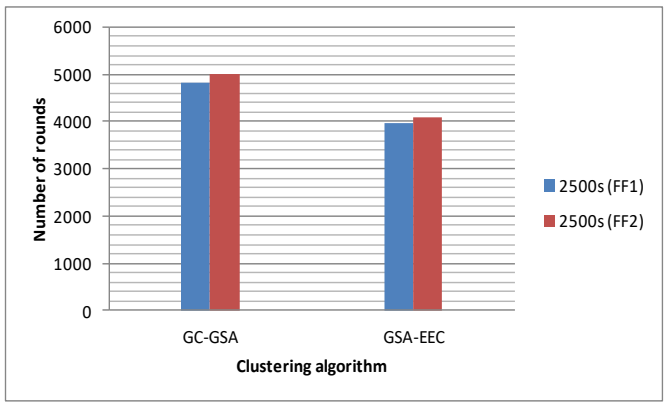

Fig. 6. Network Lifetime for Fitness Functions #1 and #2.

A relevant increase of approximately 4% is observed for the GC-GSA algorithm vs. an increase of approximately 2.5% for the GSA-EEC algorithm. As it was expected the network lifetime for both algorithms is not influenced in the same extent as the energy consumption (due to the fact that it has already reached quite high values for the specific number of sensors and the variance of the sensors' energy consumption has already been substantially restricted). As it is shown in all the measurements presented above, the proposed new fitness function leads to clearly better performance since it can represent more accurately the slight-extent modifications during the algorithm's execution and allows better balancing and accurate justification of the computed values.

### C. Evaluating the Scalability of our Combined Approach

Moreover, in order to further examine the high efficiency and scalability of our combined protocol (GC-GSA algorithm) with use of the new proposed fitness function (FF #2), we've run additional experiments for very large WSN deployments - up to 10000 sensors. Specifically, we've examined the behavior of the GC-GSA protocol (considering the energy consumption and network lifetime as the basic performance metrics) for 4000, 6000, 8000 and 10000 sensor nodes, over a progressively growing deployment areas - from 800x800$m^2$ to 2000x2000$m^2$ (i.e. 800x800$m^2$, 1200x1200 $m^2$, 1600x1600$m^2$ and 2000x2000$m^2$ terrain respectively).

Moreover, we've used as the base for comparison the network lifetime achieved for 2500 sensor nodes with the use of FF #2, which is approximately equal (see also Fig. 6 and the relevant discussion) to 5000 rounds. We've chosen progressively growing deployment areas in order to test our approach in more realistic/practical cases; note here that if the size of the deployment area had been kept the same (i.e. 480x480$m^2$ as it was for 2500 sensors) our combined approach would lead to almost equal measurements (only a slight decrease would be observed in the network lifetime) since our first-level clustering protocol is not influenced significantly (by its nature - see also [22-23] with respect to the quite similar watershed-based clustering protocol) by the increased density of sensors within the deployment area.

As it is shown in Fig. 8 there is a progressive decrease in the network lifetime when the number of the deployed sensors increase, which is more clear/significant for 10000 sensor nodes (over a 2000x2000 terrain). More concretely, the corresponding decrease for 4000, 6000, 8000 and 10000 sensors is approximately equal to 2%, 5,4%, 9.1% and 13.5% respectively. This decrease may not be considered insignificant, (at least for 8000 and 10000 sensors), however it is quite expected due to the large extent of the corresponding deployment areas. Due to the progressively growing deployment area the number of CHs and their sizes increase with a much less structured/controllable way (comparing to the case of keeping the deployment area the same), thus making much more difficult to keep the desired balance in energy consumption. Note here also that neither the GSA-EEC algorithm (GSA-based clustering algorithm only) or the WA-GSA algorithms would lead to better results (probably not even comparable) for such large numbers of sensors and deployment areas.

The observed decrease in the network lifetime is mainly caused by the relevant increase in the energy consumed by the sensor nodes (in average), as it is shown in more details in Fig. 7. In Fig. 7 one can easily observe (staring specifically at the relevant curves for 8000 and 10000 sensors) that the average energy consumption increases slightly till the execution of approximately 4000 rounds (due to the reasons referred above), and progressively more sharply afterwards, as the result of the energy exhaustion of some sensors and the end of network lifetime in every case. Overall, we can say that the proposed GC-GSA algorithm scales quite well even for very large number of sensors and deployment areas, thus making itself a promising choice for such extent realistic applications. Moreover, as it was also discussed in subsection III.D, it can be easily combined and integrated with one or more (mobile or not) sinks/collectors, thus forming a robust and flexible data gathering solution for very large-scale WSN based application environments.

### D. Comparing to other Large-Scale Clustering Approaches of the Literature

Furthermore, in order to have a more clear sense of the performance of our two-level clustering protocol we've performed additional experiments comparing the proposed approach to other clustering approaches of the literature, such as (a) TL-LEACH [26] and EEHC [27], which are two of the most known two-level clustering protocols, and (b) the protocols presented in [30] and [31] which are two of the most valuable recent clustering approaches for large-scale WSN.

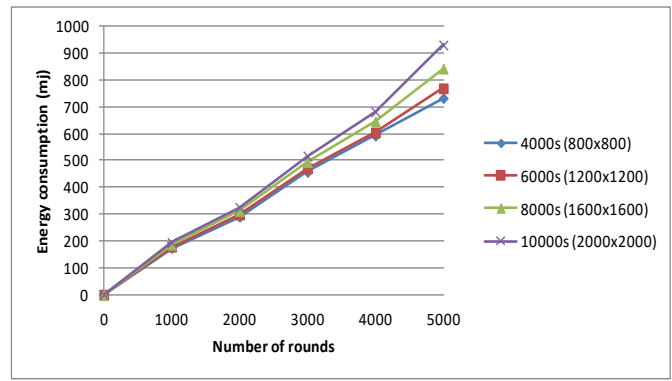

Fig. 7. Average Energy Consumption for Very Large Number of Sensors.

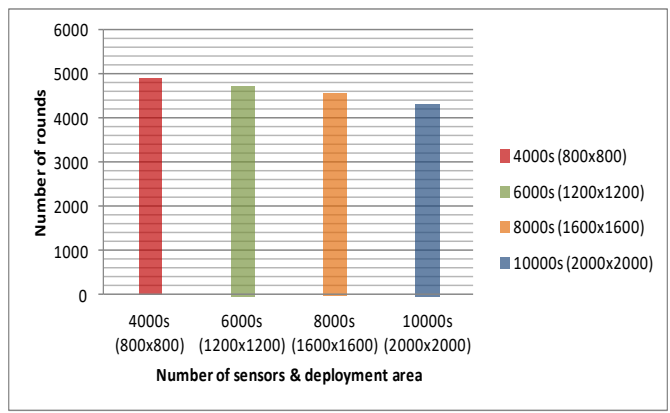

Fig. 8. Network Lifetime for Very Large Number of Sensors.

Specifically, we've run experiments for n=2500 sensors over a 480x480m2 terrain (which is a representative case of a very large WSN instance), and we've compared the performance of all the protocols over that instance. The corresponding results with regard to the energy consumption of all protocols in average are given in Fig. 9. In Fig. 10, the lifetime of the network for all the algorithms is shown (with additional measurements taken and presented for n=1500 sensors over a 360x360m2 for comparison reasons).

In [26] the authors propose a hierarchical protocol that consists of a two levels of clustering (TL-LEACH), aiming at managing the consumed energy in a more efficient way. The TL-LEACH protocol rotates both the first-level and second-level cluster-heads at random. In this way a two-level hierarchy is built, where it is possible, thus leading to a more efficient distribution of the energy reserves between the sensor nodes, which tends to be critical mainly when we have a quite dense WSN. Moreover, it is shown to perform much better than LEACH when the consumed energy and the lifetime of stand as the basic performance metrics. In [27], a random cluster-based hierarchy is proposed to organize the sensor nodes in a distributed manner. Their basic algorithm is then extended (EEHC) towards the construction of additional levels of cluster-heads, and finally a significant increase is observed with respect to the consumed energy. In [30] the authors first provide a thorough description and analysis with regard to the concept of cluster-based routing, and afterwards they introduce a corresponding combined cluster-based protocol (JCR), aiming at increased reliability and efficiency during the data gathering task in very large networks. In the JCR protocol the use of a back-off timer is followed, as well as a gradient-based protocol for routing, in order to conclude to a sufficiently connected network topology with efficient internal routing, given a specific maximum value with regard to the acceptable range of transmission. Further, in [31], the efficient solution of the 'hot spot' problem is the main objective of the authors. Their effort is based on the suitable rotation of the role of cluster-head between al the sensors, as well as adjusting appropriately the size of the formed clusters. The proposed protocol (UCF) first aims at selecting as probable cluster-heads the sensors that have more residual energy in the local area. Next, a fuzzy-logic technique is employed for the adjustment of the radius of the cluster radius. Simulation results show that the above protocol achieves significant improvements in the basic performance metrics.

As it is shown in Fig. 9 the energy consumption of the GC-GSA algorithm is less in average than all the competing protocols. The network lifetime (Fig. 10) is also steadily kept in high levels (around 5000 rounds), and its superiority over the other protocols is clear for both 1500 and 2500 sensors. The performance of TL-LEACH [26] is quite satisfactory, especially for 1500 sensors. However, although it also uses (among else) localized coordination to enable scalability, it employs a probability model for CH selection and so its energy efficiency can't be maximized. As a result it cannot preserve competitive efficiency comparing to GC-GSA for large and very large number of sensors. The EEHC protocol has similar behaviour with TL-LEACH since it's a randomized approach too; Moreover it performs slightly better due to its modular nature and the fact that it pays more attention in energy efficiency than TL-LEACH (stochastic geometry techniques are also used to improve the energy consumption). However for the same reasons it's also not highly competitive comparing to GC-GSA for large and very number of sensors.

The best performance among the other protocols is given by the UCF protocol [31], which leads to similar (slightly worse) network lifetime with GC-GSA and average network consumption, by combining unequal clustering and residual energy based CH selection with fuzzy logic to adjust the cluster radius. The fact that no other factors than the residual energy are taken in account for CH selection in each region, as well as the uncertainty nature of the fuzzy logic procedure makes UCF not scaling the same well for very large numbers of sensors (i.e. for 1500 and 2500 sensors in our case) as for smaller in size networks.

The JCR protocol [30] also performs very well for large networks since it selects a set of cluster-heads that helps (a) to organize a quite balanced cluster hierarchy (taking in account multiple factors for the procedure of CH selection), and (b) to construct a backbone of good connectivity for internal cluster-based routing. However it also doesn't scale the same well for very large numbers of sensors (and deployment areas), mainly because (as a result of the increase in the network size) the residual energy factor gets less critical (than it should be) on the computation of the back-off time of each node.

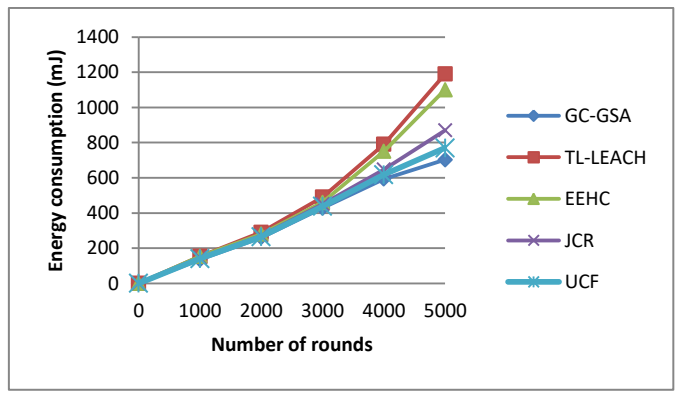

Fig. 9. Average Energy Consumption for GC-GSA, TL-LEACH, EEHC, JCR amd UCF (for 2500 Sensors).

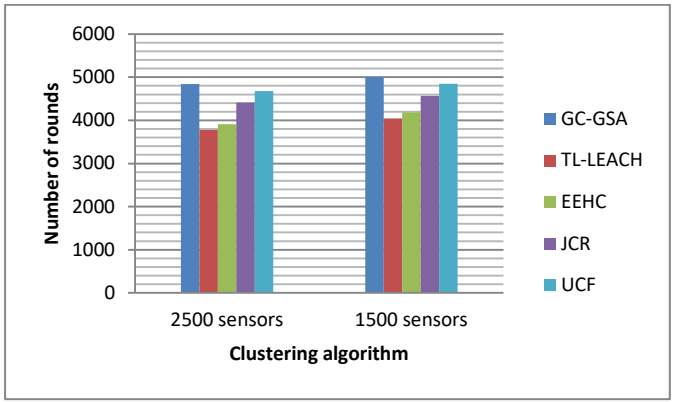

Fig. 10. Network Lifetime for GC-GSA, TL-LEACH, EEHC, JCR and UCF (for 1500 and 2500 Sensors).

### E. Evaluating the First-Level Clustering Algorithm

Finally, in Table II and Fig. 11, 12 we've performed a basic set of experiments to show the worth of using the proposed modified first-level clustering algorithm instead of the more complex watershed-based algorithm given in [22-23]. More concretely, supposing that we use the algorithm of [22-23] as the first-level clustering algorithm in our combined approach, and keeping the GSA oriented algorithm as the cluster formation protocol of the second level, we conclude to a related alternative combined approach, i.e. WA-GSA. In the above context, we've run some of the experiments introduced in Section 4.1 (for n=1500 and n=2500 sensors over 360x360m2 and 480x480m2 terrains, respectively), aiming at a fair performance comparison of the two algorithms in large-scale WSN instances. For both algorithms (GC-GSA and WA-GSA) we've used the new proposed fitness function (FF #2) in order to have optimized performance behaviour.

As it is shown in Fig. 11 the average energy consumption of the WA-GSA algorithm is slightly larger than the one of the GC-GSA algorithm. Specifically, a slight increase is observed which ranges from 2.5% to 4% (depending on the number of rounds) for n=2500 sensors and from 2% to 3% for n=1500 sensors. Actually, in the beginning the behavior of both algorithms is approximately the same, whereas in the progress of the execution the WA-GSA algorithm tends to spend slightly more energy, which becomes more clear for larger number of sensors (n=2500). The latter is naturally expected, since with larger number of sensors we conclude to larger number of CHs in general (for both algorithms); and more concretely to larger differences in the number of cluster-heads between each other, which influences even more significantly the behaviour of the WA-GSA algorithm during its execution.

Furthermore, in Fig. 12 the corresponding difference in network lifetime is shown between the two algorithms. Here, one may notice also a slight decrease in the network lifetime for the WA-GSA algorithm, which is approximately equal to 2.5% and 1.5% for n=2500 and n=1500 respectively (the network lifetime is influenced in less extent for the same reasons as discussed for Fig. 6). As a general conclusion, the corresponding differences are not quite significant; however for large and very large number of sensors they make the choice of GC-GSA algorithm clearly preferable. As it was also discussed in Section 3.1 the slightly superior performance of our first-level clustering protocol against the watershed-based one of [22-23] is due to the less number of CHs it leads (which is a crucial parameter in the context of the present gateways-based WSN environment), whereas also the execution of the cluster formation procedure is quite faster. More concretely, the exact number of clusters (and CHs) produced for both algorithms in each set of experiments are given in Table II. Instead, in the context of other WSN environments (e.g. data gathering applications with a mobile collector, as in [22-23]) the performance of the watershed-based clustering protocol should be the preferable one.

TABLE II. NUMBER AND SIZE OF CLUSTERS FOR GC-GSA AND WA-GSA

| Sensors | # of Clusters (WA-GSA) | Nodes per Cluster (WA-GSA) | # of Clusters (GC-GSA) | Nodes per Cluster (GC-GSA) |
|---|---|---|---|---|
| 500 | 105 | 4.8 | 94 | 5.3 |
| 1500 | 133 | 11.3 | 120 | 12.5 |
| 2500 | 187 | 13.4 | 171 | 14.6 |
| 4000 | 251 | 15.9 | 231 | 17.3 |
| 6000 | 321 | 18.7 | 295 | 20.3 |
| 8000 | 408 | 19.6 | 368 | 21.7 |
| 10000 | 495 | 20.2 | 447 | 22.4 |

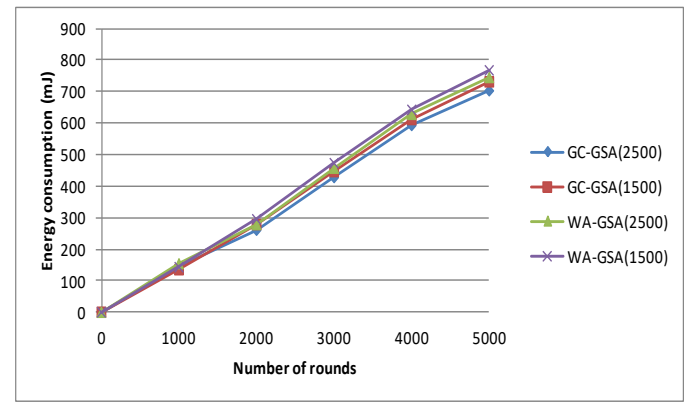

Fig. 11. Average Energy Consumption for GC-GSA and WA-GSA.

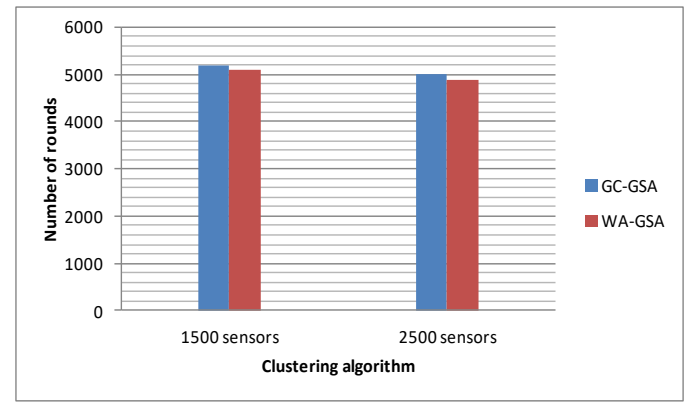

Fig. 12. Network Lifetime for GC-GSA and WA-GSA.

## V. Conclusions

Throughout this paper the worth of using a novel hybrid clustering scheme that completes in two separate / hierarchical phases is demonstrated. It brings together a typical gradient protocol for clusters formation with an evolutionary optimization method that is mainly based on the recently introduced (the last decade) GSA algorithm, and trying to overcome its shortcomings on large networks. As a consequence it aims at better behaviour over large in size networks, where classical schemes in most cases lead to non-efficient solutions. The presented approach first constructs multihop clusters of evenly distributed energy reserves, in which the sensor nodes energy gets larger as coming closer to the cluster head. Thus, since the sensors near to the cluster head are more exhausted because of data forwarding, keeping the residual energy of those sensors high guarantees the prolonged and seamless operation of the WSN. Next, a suitable protocol based on the GSA runs to associate sets of cluster heads to specific gateway nodes for the eventual relaying of data to the base station. A fitness function was appropriately chosen considering both the distance from the cluster heads to the gateway nodes and the remaining energy of the gateway nodes. It was also further modified-optimized in order to gain more accurate results for large WSN instances (by providing the means for more discrete weighting of the overall formula). Extended simulated experiments have been completed to show the high efficiency and scalability of the proposed combined clustering scheme over very large WSNs, as well as its superiority over other clustering approaches of the literature. Additional sets of experiments demonstrate also separately the suitability of the proposed new fitness function against the previously existing one, as well as the worth of using the proposed modified first-level clustering algorithm against the use of other alternatives.